\documentclass[pre,aps,preprint,showpacs,preprintnumbers,amsmath,amssymb,secnumroman,eqsecnum]{revtex4}

\usepackage{graphicx}% Include figure files
\usepackage{dcolumn}% Align table columns on decimal point
\usepackage{bm}% bold math

\newcommand{\beq}{\begin{equation}}
\newcommand{\eeq}{\end{equation}}
\newcommand{\beqa}{\begin{eqnarray}}
\newcommand{\eeqa}{\end{eqnarray}}
\newcommand{\vc}[1]{\mbox{\boldmath $#1$}}

\newcommand{\vol}[1]{{\bf #1}}

\newcommand{\du}[1]{{\bf\sf #1}}

\begin{document}

%\preprint{APS/123-QED}

\title{Hydrodynamic interactions between a sphere and a number of small particles}% Force line breaks with \\

\author{Maria L. Ekiel-Je\.zewska}

\email{mekiel@ippt.pan.pl}
\affiliation{Department of Mechanics and Physics of Fluids\\Institute of Fundamental Technological Research\\ Polish Academy of Sciences\\
Pawinskiegop 5B\\ 02-106 Warsaw\\ Poland}%
\author{B. U. Felderhof}

 \email{ufelder@physik.rwth-aachen.de}
\affiliation{Institut f\"ur Theorie der Statistischen Physik \\ RWTH Aachen University\\
Templergraben 55\\52056 Aachen\\ Germany\\
}%

\date{\today}% It is always \today, today,

\begin{abstract}
Exact expressions are derived for the pair and three-body hydrodynamic interactions between a sphere and a number of small particles immersed in a viscous incompressible fluid. The analysis is based on the Stokes equations of low Reynolds number hydrodynamics. The results follow by a combination of the solutions for flow about a sphere with no-slip boundary condition derived by Stokes and Kirchhoff, and the result derived by Oseen for the Green tensor of Stokes' equations in the presence of a fixed sphere.
\end{abstract}

\maketitle
\section{\label{I}Introduction}

The calculation of hydrodynamic interactions between spheres suspended in a viscous incompressible fluid is notoriously difficult \cite{1},\cite{2}. Even the calculation of the pair hydrodynamic interaction between two rigid spheres with no-slip boundary condition is elaborate \cite{3},\cite{4}. For more than two spheres many-body hydrodynamic interactions appear \cite{5}, and recourse must be taken to multipole expansion \cite{6}. Although an efficient numerical scheme has been developed for the accurate evaluation of hydrodynamic interactions between spheres or droplets \cite{7},\cite{8}, it remains of interest to have analytic expressions available for use in particular situations.

In the following we consider hydrodynamic interactions between a single large sphere with no-slip boundary condition and a number of small particles which can be treated in force monopole approximation. We are therefore dealing with a sphere-bead model. We use the exact solutions for flow about a sphere acted on by a force or torque, due to Stokes and Kirchhoff, respectively, as well as the exact Green tensor of the Stokes equations for a single fixed sphere, as derived by Oseen \cite{9}. Combination of these results allows the derivation of pair and three-body hydrodynamic interaction functions which yields an approximate expression for the translational mobility matrix of sphere and particles. The analytic expression becomes exact in the limit where the radii of the small particles are much less than the radius of the sphere, the separation from the sphere, and their mutual distances. In our derivation we consider a sphere and at most two small particles, but the result can be immediately generalized to a larger number of beads.

The important result of Oseen \cite{9} was first used in the context of the theory of swimming by Higdon \cite{10},\cite{11}, who studied the motion of a microorganism with a spherical cell body propelled by a cylindrical flagellum. Golestanian used Oseen's result in a sphere-bead model consisting of a sphere and two collinear beads in longitudinal motion along the common axis \cite{12}. He suggested that in the swimming motion of the linear chain structure the force on the sphere is zero. However, we show in the following that under this condition the swimming velocity vanishes. The argument is similar to that used in the derivation of the scallop theorem \cite{13}.

In a second application we consider the motion of two particles and a sphere subject to gravity and lined up either vertically or horizontally. We expand the velocities of the three bodies in inverse powers of the radius of the sphere and derive approximate results for the velocities of the three bodies. The main result is that to lowest order the sphere and the two particles move together.

The approximate results for hydrodynamic interactions of a sphere and small particles are derived in Sec. II. The swimming of a linear chain structure consisting of a sphere and two beads, as studied by Golestanian \cite{12}, is discussed in Sec. III. The sedimentation of a sphere and two collinear particles is studied in Sec. IV. In the concluding section we discuss briefly further possible applications.

\section{\label{II}Hydrodynamic interactions in small particle approximation}

In the following we shall consider an $N+1$-body system consisting of a sphere of radius $b$, labeled $0$, and $N$ much smaller particles of radii $a_1,...,a_N$, all immersed in a viscous incompressible fluid of shear viscosity $\eta$. The fluid is of infinite extent in all directions. At low Reynolds
number and on a slow time scale the flow velocity
$\vc{v}$ and the pressure $p$ satisfy the
Stokes equations \cite{1}
\begin{equation}
\label{2.1}\eta\nabla^2\vc{v}-\nabla p=0,\qquad\nabla\cdot\vc{v}=0.
\end{equation}
The flow velocity $\vc{v}$ is assumed to satisfy the no-slip boundary condition on the surface of the sphere and of the particles. Moreover, all bodies are assumed to be rotating freely, so that the torques vanish. The effect of the particles on the fluid will be treated in force monopole approximation.

In principle the hydrodynamic interactions are embodied in the $(3N+3)\times(3N+3)$  mobility matrix $\vc{\mu}$ which relates the translational velocities $(\vc{U}_0,...,\vc{U}_N)$ to the $N+1$ forces $(\vc{F}_0,...,\vc{F}_N)$ exerted on the bodies via
 \begin{equation}
\label{2.2}\vc{U}_j=\sum^N_{k=0}\vc{\mu}_{jk}\cdot\vc{F}_k,\qquad (j=0,...,N),
\end{equation}
 where each pair term $\vc{\mu}_{jk}$ depends on the positions $(\vc{R}_0,...,\vc{R}_N)$ of all centers. We shall derive approximate simplified expressions for the mobility tensors, based on the assumption that the particles make their presence felt only via the force they exert on the fluid and otherwise are carried along by the flow caused by the sphere and the other particles. In the approximation the element $\vc{\mu}_{00}$ is independent of position, elements $\vc{\mu}_{0j}$ and $\vc{\mu}_{j0}$ with $j>0$ depend only on $\vc{R}_j-\vc{R}_0$ via a pair hydrodynamic interaction, and elements $\vc{\mu}_{jk}$, with $j>0,k>0$ depend only on $\vc{R}_j-\vc{R}_0$ and $\vc{R}_k-\vc{R}_0$ via a three-body hydrodynamic interaction. In order to derive the expression for the pair hydrodynamic interactions $\vc{\mu}_{0j}$ and $\vc{\mu}_{j0}$ it suffices to consider the sphere and a single small particle. In order to derive the three-body hydrodynamic interaction we must consider the sphere and two small particles.

 We use Cartesian coordinates with origin at the center of the sphere. We recall first that in the absence of particles, the sphere moving with translational velocity $\vc{U}_0$ and rotational velocity $\vc{\Omega}_0$ would exert a force $\vc{\mathcal{F}}_0$ and a torque $\vc{\mathcal{T}}_0$ on the fluid given by
\begin{equation}
\label{2.3}\vc{\mathcal{F}}_0=6\pi\eta b\vc{U}_0,\qquad\vc{\mathcal{T}}_0=8\pi\eta b^3\vc{\Omega}_0,
\end{equation}
according to Stokes and Kirchhoff, respectively. The corresponding flow pattern is given by
\begin{equation}
\label{2.4}\vc{v}^{SK}(\vc{r})=\frac{1}{8\pi\eta}\bigg[\frac{\du{I}+\hat{\vc{r}}\hat{\vc{r}}}{r}+\frac{b^2}{3}\frac{\du{I}-3\hat{\vc{r}}\hat{\vc{r}}}{r^3}\bigg]\cdot\vc{\mathcal{F}}_0
-\frac{1}{8\pi\eta}\frac{\vc{r}}{r^3}\times\vc{\mathcal{T}}_0,
\end{equation}
where $\du{I}$ is the unit tensor. It is convenient to use the first part in Eq. (2.4) to define the Stokes tensor field
\begin{equation}
\label{2.5}\du{M}^{St}_t(\vc{r})=\frac{1}{8\pi\eta}\bigg[\frac{\du{I}+\hat{\vc{r}}\hat{\vc{r}}}{r}+\frac{b^2}{3}\frac{\du{I}-3\hat{\vc{r}}\hat{\vc{r}}}{r^3}\bigg].
\end{equation}

Next we consider a sphere of radius $b$ and a single small particle, both immersed in a viscous
fluid. The pair hydrodynamic interaction between particle and sphere is embodied in the six-dimensional translational mobility matrix relating the sphere velocity $\vc{U}_0$ and the particle velocity $\vc{U}_1$ to the forces $\vc{F}_0$ and $\vc{F_1}$ acting on sphere and particle according to
\begin{eqnarray}
\label{2.6}\vc{U}_0&=&\vc{\mu}_{00}\cdot\vc{F}_0+\vc{\mu}_{01}\cdot\vc{F}_1,\nonumber\\
\vc{U}_1&=&\vc{\mu}_{10}\cdot\vc{F}_0+\vc{\mu}_{11}\cdot\vc{F}_1.
\end{eqnarray}
We consider first the situation with $\vc{F}_1=0$ with a force $\vc{F}_0$ acting on the sphere.
In our approximation the particle is carried along by the flow pattern generated by $\vc{F}_0$, so that the parts $\vc{\mu}_{00}$ and $\vc{\mu}_{10}$ of the mobility matrix are given simply by
 \begin{equation}
\label{2.7}\vc{\mu}_{00}=\frac{1}{6\pi\eta b}\;\du{I},\qquad\vc{\mu}_{10}=\du{M}^{St}_t(\vc{r}_1),
\end{equation}
with relative vector $\vc{r}_1=\vc{R}_1-\vc{R}_0$.

In order to derive approximate expressions for the parts $\vc{\mu}_{01}$ and $\vc{\mu}_{11}$ of the mobility matrix in Eq. (2.6) we consider a flow situation where a force $\vc{F}_1$ is applied to the particle, but the sphere is moving freely with translational velocity $\vc{U}_0$ and rotational velocity $\vc{\Omega}_0$ such that it exerts no force or torque on the fluid. First we recall an important result derived by Oseen \cite{9}.

For a fixed sphere centered at the origin and with a point force $\vc{F}_1$ acting on the fluid at $\vc{r}_1$ Oseen derived the flow pattern \cite{9}
\begin{equation}
\label{2.8}\vc{v}^{Os}(\vc{r})=\du{T}(\vc{r},\vc{r}_1)\cdot\vc{F}_1
\end{equation}
with Green tensor given by
\begin{equation}
\label{2.9}\du{T}(\vc{r},\vc{r}_1)=\du{T}_0(\vc{r}-\vc{r}_1)+\du{T}_R(\vc{r},\vc{r}_1),
\end{equation}
where $\du{T}_0(\vc{r}-\vc{r}_1)$ is the tensor for infinite space in the absence of the sphere, given by
\begin{equation}
\label{2.10}\du{T}_0(\vc{r})=\frac{1}{8\pi\eta}\frac{\du{I}+\hat{\vc{r}}\hat{\vc{r}}}{r},
\end{equation}
and $\du{T}_R(\vc{r},\vc{r}_1)$ is the reflection tensor
\begin{eqnarray}
\label{2.11}\du{T}_R(\vc{r},\vc{r}_1)&=&\frac{-1}{8\pi\eta}\bigg[\frac{b}{r_1\overline{d}}\;\du{I}+\frac{b^3}{r_1^3\overline{d}\;^3}(\vc{r}-\overline{\vc{r}_1})(\vc{r}-\overline{\vc{r}_1})\nonumber\\
&+&\frac{r_1^2-b^2}{r_1}\bigg(\frac{1}{b^3\overline{d}}\overline{\vc{r}_1}\;\overline{\vc{r}_1}
-\frac{b}{r_1^2\overline{d}\;^3}\big(\overline{\vc{r}_1}(\vc{r}-\overline{\vc{r}_1})+(\vc{r}-\overline{\vc{r}_1})\overline{\vc{r}_1}\big)\nonumber\\
&-&\frac{2}{b^3}\overline{\vc{r}_1}\;\overline{\vc{r}_1}\;\overline{\vc{r}_1}\cdot\nabla\frac{1}{\overline{d}}\bigg)+(r^2-b^2)\nabla\vc{\phi}\bigg],
\end{eqnarray}
where the image point $\overline{\vc{r}_1}$ is defined by
\begin{equation}
\label{2.12}\overline{\vc{r}_1}=\frac{b^2}{r_1^2}\vc{r}_1,
\end{equation}
and $\overline{d}$ is the distance from the field point $\vc{r}$ to the image point
 \begin{equation}
\label{2.13}\overline{d}=|\vc{r}-\overline{\vc{r}_1}|.
\end{equation}
Furthermore, the vector function $\vc{\phi}$ is given by
\begin{eqnarray}
\label{2.14}\vc{\phi}(\vc{r},\vc{r}_1)&=&\frac{r_1^2-b^2}{2r_1^3}\bigg(\frac{3}{b\overline{d}}\;\vc{r}_1+\frac{b}{\overline{d}^3}\;(\vc{r}-\overline{\vc{r}_1})
+\frac{2}{b}\;\vc{r}_1\;\overline{\vc{r}_1}\cdot\nabla\frac{1}{\overline{d}}\nonumber\\
&+&\frac{3b}{\overline{r}_1}\frac{\partial}{\partial \overline{\vc{r}}_1}\log\frac{\overline{r}_1\overline{d}+\vc{r}\cdot\overline{\vc{r}}_1-\overline{r}_1^2}{b\overline{r}_1+\vc{r}\cdot\overline{\vc{r}}_1}\bigg).
\end{eqnarray}
The tensor $\du{T}(\vc{r},\vc{r}_1)$ vanishes for points $\vc{r}$ located on the surface of the sphere, so that the no-slip boundary condition for a fixed sphere is satisfied.

From the asymptotic behavior of the reflection tensor $\du{T}_R(\vc{r},\vc{r}_1)$ for large $r$ we see that in Oseen's Green function the sphere exerts a force $\vc{\mathcal{F}}$ and a torque $\vc{\mathcal{T}}$ on the fluid given by
\begin{equation}
\label{2.15}\vc{\mathcal{F}}=-6\pi\eta b\du{M}^{St}_t(\vc{r}_1)\cdot\vc{F}_1,\qquad\vc{\mathcal{T}}=-\frac{b^3}{r_1^3}\;\vc{r}_1\times\vc{F}_1.
\end{equation}
We must subtract the Stokes-Kirchhoff flow generated by the force $\vc{\mathcal{F}}$ and the torque $\vc{\mathcal{T}}$ from Oseen's solution in order to get the flow pattern for a freely moving sphere.

Alternatively the expressions in Eq. (2.15) can be derived by use of Fax\'en's theorems applied to the flow $\du{T}_0(\vc{r}-\vc{r}_1)\cdot\vc{F}_1$ generated by a force $\vc{F}_1$ at $\vc{r}_1$ in infinite fluid in the absence of the sphere \cite{1}. The expressions can be written as
\begin{equation}
\label{2.16}\vc{\mathcal{F}}=\vc{\rho}^{tt}(\vc{r}_1)\cdot\vc{F}_1,\qquad\vc{\mathcal{T}}=\vc{\rho}^{rt}(\vc{r}_1)\cdot\vc{F}_1,
\end{equation}
with reflection tensors given by
\begin{equation}
\label{2.17}\vc{\rho}^{tt}(\vc{r}_1)=-6\pi\eta b\du{M}^{St}_t(\vc{r}_1),\qquad\rho^{rt}_{\alpha\beta}(\vc{r}_1)=\frac{b^3}{r_1^3}\;\varepsilon_{\alpha\beta\gamma}r_{1\gamma},
\end{equation}
where $\varepsilon_{\alpha\beta\gamma}$ is the Levi-Civita tensor. The flow generated by the force $\vc{F}_1$ in the presence of the freely moving sphere can be expressed as
\begin{equation}
\label{2.18}\hat{\vc{v}}^{Os}(\vc{r})=\hat{\du{T}}(\vc{r},\vc{r}_1)\cdot\vc{F}_1,
\end{equation}
with modified Oseen tensor
\begin{equation}
\label{2.19}\hat{\du{T}}(\vc{r},\vc{r}_1)=\du{T}(\vc{r},\vc{r}_1)+\du{V}(\vc{r},\vc{r}_1),
\end{equation}
with tensor function $\du{V}(\vc{r},\vc{r}_1)$ given by
\begin{equation}
\label{2.20}\du{V}(\vc{r},\vc{r}_1)=\du{V}^t(\vc{r},\vc{r}_1)+\du{V}^r(\vc{r},\vc{r}_1),
\end{equation}
with translational part
\begin{equation}
\label{2.21}\du{V}^t(\vc{r},\vc{r}_1)=-\du{M}^{St}_t(\vc{r})\cdot\vc{\rho}^{tt}(\vc{r}_1)=6\pi\eta b\du{M}^{St}_t(\vc{r})\cdot\du{M}^{St}_t(\vc{r}_1),
\end{equation}
and rotational part
\begin{equation}
\label{2.22}\du{V}^r(\vc{r},\vc{r}_1)=-\frac{b^3}{8\pi\eta r^3r^3_1}\;\big[\vc{r}_1\vc{r}-\vc{r}_1\cdot\vc{r}\;\du{I}\big].
\end{equation}

In order to derive the approximate expression for the mobility tensor $\vc{\mu}_{01}$ in Eq. (2.6) we consider the modified Oseen  flow $\hat{\vc{v}}^{Os}(\vc{r})$ on the surface of the sphere. We denote a point on the surface of the sphere by $\vc{s}$ and note from Eq. (2.5) that
\begin{equation}
\label{2.23}\du{M}^{St}_t(\vc{s})=\frac{1}{6\pi\eta b}\;\du{I}.
\end{equation}
Using also $\du{T}(\vc{s},\vc{r}_1)=\du{0}$ we find from the translational part of the flow $\hat{\vc{v}}^{Os}(\vc{r})$ on the surface of the sphere
\begin{equation}
\label{2.24}\vc{\mu}_{01}=\du{M}^{St}_t(\vc{r}_1),
\end{equation}
so that $\vc{\mu}_{01}=\vc{\mu}_{10}$ by comparison with Eq. (2.7), in agreement with the symmetry of the mobility matrix which holds on general grounds.

We derive the approximate expression for the mobility tensor $\vc{\mu}_{11}$ by considering the reflected part of the modified Oseen flow $\hat{\vc{v}}^{Os}(\vc{r})$ at the point $\vc{r}=\vc{r}_1$. This yields
\begin{equation}
\label{2.25}\vc{\mu}_{11}=\frac{1}{6\pi\eta a_1}\;\du{I}+\hat{\du{T}}_R(\vc{r}_1,\vc{r}_1),
\end{equation}
where $a_1$ is the particle radius.
From Eqs. (2.11) and (2.19) we find the explicit expression
\begin{equation}
\label{2.26}\hat{\du{T}}_R(\vc{r}_1,\vc{r}_1)=\frac{-1}{8\pi\eta}\bigg[\frac{b^3}{3r_1^6}\frac{15r_1^4-7b^2r_1^2
+b^4}{r_1^2-b^2}\;\hat{\vc{r}}_1\hat{\vc{r}}_1+\frac{b^5}{12r_1^6}\frac{17r_1^2+b^2}{r_1^2-b^2}\;(\du{I}-\hat{\vc{r}}_1\hat{\vc{r}}_1)\bigg].
\end{equation}
Note that this becomes singular at short distance $r_1-b$ and shows a rapid decay at large distance. The tensor $\vc{\mu}_{11}$ is obviously symmetric.

We can use the same formalism to derive the approximate expression for the mobility tensor $\vc{\mu}_{jk}$, with $j>0,k>0$. It suffices to consider a sphere and two small particles located at $\vc{R}_1$ and $\vc{R}_2$. The linear relation between velocities and spheres becomes
\begin{eqnarray}
\label{2.27}\vc{U}_0&=&\vc{\mu}_{00}\cdot\vc{F}_0+\vc{\mu}_{01}\cdot\vc{F}_1+\vc{\mu}_{02}\cdot\vc{F}_2,\nonumber\\
\vc{U}_1&=&\vc{\mu}_{10}\cdot\vc{F}_0+\vc{\mu}_{11}\cdot\vc{F}_1+\vc{\mu}_{12}\cdot\vc{F}_2,\nonumber\\
\vc{U}_2&=&\vc{\mu}_{20}\cdot\vc{F}_0+\vc{\mu}_{21}\cdot\vc{F}_1+\vc{\mu}_{22}\cdot\vc{F}_2.
\end{eqnarray}
Most parts of the mobility matrix are given by expressions derived above. Only the parts $\vc{\mu}_{12}$ and $\vc{\mu}_{21}$ require further consideration.
In generalization of Eq. (2.25) we have
\begin{equation}
\label{2.28}\vc{\mu}_{12}=\hat{\du{T}}(\vc{r}_1,\vc{r}_2),\qquad\vc{\mu}_{21}=\hat{\du{T}}(\vc{r}_2,\vc{r}_1),
\end{equation}
with $\vc{r}_1=\vc{R}_1-\vc{R_0}$ and $\vc{r}_2=\vc{R}_2-\vc{R_0}$.
The symmetry relation
\begin{equation}
\label{2.29}\vc{\mu}_{21}=\tilde{\vc{\mu}}_{12}
\end{equation}
is satisfied, in agreement with general arguments.

The expression for $\vc{\mu}_{12}$ given by Eq. (2.28) is complicated, but it simplifies for configurations for which $\vc{r}_1$ and $\vc{r}_2$ are collinear with the origin, so that $\hat{\vc{r}}_1=\hat{\vc{r}}_2$. For such configurations we find
\begin{eqnarray}
\label{2.30}\vc{\mu}_{12}&=&\du{T}_0(\vc{r}_1-\vc{r}_2)+\frac{b}{8\pi\eta}\bigg[\bigg(\frac{(3r_1^2-b^2)(3r_2^2-b^2)}{3r_1^3r_2^3}-
\frac{3r_1^2r_2^2-b^2(r_1^2+4r_1r_2+r_2^2)+3b^4}{(r_1r_2-b^2)^3}\bigg)\hat{\vc{r}}_1\hat{\vc{r}}_1\nonumber\\
&-&\frac{b^4}{12r_1^3r_2^3}\frac{17r_1^3r_2^3-3b^2r_1r_2(3r_1^2+5r_1r_2+3r_2^2)+3b^4(r_1^2+3r_1r_2+r_2^2)+b^6}{(r_1r_2-b^2)^3}(\du{I}-\hat{\vc{r}}_1\hat{\vc{r}}_1)\bigg].\nonumber\\
\end{eqnarray}
The first term is the pair hydrodynamic interaction between two small particles.
The second term depends on both $\vc{r}_1=\vc{R}_1-\vc{R}_0$ and $\vc{r}_2=\vc{R}_2-\vc{R}_0$, and it therefore represents a three-body hydrodynamic interaction. Note that the interaction becomes singular as both $r_1$ and $r_2$ tend to the sphere radius $b$. It decays as $1/(r_1r_2)^2$ as both $r_1$ and $r_2$ tend to infinity.

The expressions we have derived above can be used as approximations in the many-body mobility matrix in Eq. (2.2). If we regard $a$ as a typical small particle radius, $d$ as a typical distance between small particles, and $h$ as the minimum separation distance of the center of a small particle from the surface of the sphere, then the ratios $a/b$, $a/d$ and $a/h$ may be regarded as small parameters. It follows from a consideration of the multipole expansion of the exact mobility matrix that our expressions represent the first few terms in a systematic expansion in powers of the small parameters. We call our expressions the small particle approximation to the mobility matrix. In the following we consider several applications of our results.

 \section{\label{III}Application to swimming with a cargo}

 The present work was inspired by a derivation of Golestanian \cite{12}, who considered swimming of a linear chain structure consisting of a sphere and two small particles with periodic motions of the three bodies along a common axis. The derivation generalized earlier work of Golestanian et al. \cite{14},\cite{15} on the swimming of a linear structure of three small particles. In his derivation Golestanian \cite{12} suggested that the force on the sphere vanishes, so that the swimming could be interpreted as the pushing of a cargo by two small particles. We show below that for vanishing force $\vc{F}_0$ the structure does not swim.

 We investigate whether the three-body structure can swim with just $\vc{F}_1(t)$ and $\vc{F}_2(t)$ as actuating forces. The forces are assumed to vary periodically in time with period $T$. In swimming the total force must vanish, so that with the cargo condition $\vc{F}_0=0$ we must require $\vc{F}_2(t)=-\vc{F}_1(t)$ at any time $t$. We can consider motion along the $x$ axis, so that $\vc{F}_1(t)=F_1(t)\vc{e}_x$, $\vc{F}_2(t)=-F_1(t)\vc{e}_x$. The relative positions are $\vc{r}_1(t)=(x_1(t)-x_0(t),0,0)$ and $\vc{r}_2(t)=(x_2(t)-x_0(t),0,0)$. The equations of Stokesian dynamics which follow from Eq. (2.27) reduce to
\begin{eqnarray}
\label{3.1}\frac{dx_0}{dt}&=&(\mu_{01}-\mu_{02})F_1,\nonumber\\
\frac{dx_1}{dt}&=&(\mu_{11}-\mu_{12})F_1,\nonumber\\
\frac{dx_2}{dt}&=&(\mu_{12}-\mu_{22})F_1,
\end{eqnarray}
with scalar mobility coefficients $\mu_{ij}$ which depend on the relative coordinates $r_1=x_1-x_0$ and $r_2=x_2-x_0$. The equations of motion for the relative coordinates are
\begin{eqnarray}
\label{3.2}
\frac{dr_1}{dt}&=&(\mu_{11}-\mu_{12}-\mu_{01}+\mu_{02})F_1,\nonumber\\
\frac{dr_2}{dt}&=&(\mu_{12}-\mu_{22}-\mu_{01}+\mu_{02})F_1.
\end{eqnarray}
Eliminating $F_1$ and $dt$ we obtain the set of ordinary differential equations
\begin{equation}
\label{3.3}
\frac{dr_2}{dr_1}=A(r_1,r_2),\qquad
\frac{dx_0}{dr_1}=B(r_1,r_2),
\end{equation}
with functions
\begin{eqnarray}
\label{3.4}A(r_1,r_2)&=&\frac{\mu_{12}-\mu_{22}-\mu_{01}+\mu_{02}}{\mu_{11}-\mu_{12}-\mu_{01}+\mu_{02}},\nonumber\\
B(r_1,r_2)&=&\frac{\mu_{01}-\mu_{02}}{\mu_{11}-\mu_{12}-\mu_{01}+\mu_{02}}.
\end{eqnarray}
From Eq. (3.3) we can obtain $r_2$ and $x_0$ as functions of $r_1$ and initial values $(r_{10},r_{20},x_{00})$. On the other hand, $r_1(t)$ is obtained as a periodic function of time from Eq. (3.2). Therefore $x_0(t)$ is also a periodic function of time, and there is no net swimming displacement. The argument is independent of the precise dependence of the mobility functions on relative coordinates found in the preceding section.

 \section{\label{IV}Sedimentation of three spheres}

 As a second application we consider sedimentation of three collinear spheres, either lined up vertically or horizontally. We consider two small spheres labeled $1,2$ of equal radius $a$  and a big sphere of radius $b$, labeled $0$. The center of the first small sphere is at separation distance $h$ from the surface of the big sphere, followed by the second one at center-to-center distance $d$ from the first one. The force of gravity acting on the big sphere along the vertical is $F_0$, and for the same mass density the forces acting on the small spheres are $F_1=F_2=(a^3/b^3)F_0$. We study the sphere velocities for small ratios $a/b,\;h/b$ and $d/b$, as well as small $a/h$ and $a/d$. Thus the particles are relatively close to the sphere, and the point particle approximation does not apply. The main result is that to lowest order the sphere and the particles move together at the same speed, an effect not predicted by the point particle approximation, but found also in Rotne-Prager approximation \cite{16}.

 Consider first the three spheres lined up along the vertical. To lowest order in $1/b$ the three spheres move together at the same velocity $U_{00}=F_0/(6\pi\eta b)$, given by Stokes' law for the big sphere. The first few correction terms to the velocity of the big sphere in an expansion in powers of $1/b$ are given by
 \begin{eqnarray}
\label{4.1}\frac{U_0}{U_{00}}=1+2\frac{a^3}{b^3}+O(b^{-5}).
\end{eqnarray}
The first two terms correspond to Stokes' law for the total force $F_0+F_1+F_2$. The lowest order correction terms to the velocity of the first small sphere are given by
 \begin{equation}
\label{4.2}\frac{U_1}{U_{00}}=1-\frac{3h^2+O(a^2)}{2b^2}+O(b^{-3}).
\end{equation}
The lowest order correction terms to the velocity of the second small sphere are given by
 \begin{equation}
\label{4.3}\frac{U_2}{U_{00}}=1-\frac{3(h+d)^2+O(a^2)}{2b^2}+O(b^{-3}).
\end{equation}
Hence the velocities of both spheres are less than $U_{00}$, the second slightly more than the first. The two small particles lag behind the big sphere.

Next consider the three spheres lined up along the horizontal. To lowest order in $1/b$ the three spheres move together at the same velocity $U_{00}=F_0/(6\pi\eta b)$, given by Stokes' law for the big sphere. The first few correction terms to the velocity of the big sphere are given by
 \begin{eqnarray}
\label{4.4}\frac{U_0}{U_{00}}=1+2\frac{a^3}{b^3}-\frac{3}{2}(2h+d)\frac{a^3}{b^4}+O(b^{-5}).
\end{eqnarray}
The first two terms correspond to Stokes' law for the total force $F_0+F_1+F_2$. The first few correction terms to the velocity of the first small sphere are given by
 \begin{equation}
\label{4.5}\frac{U_1}{U_{00}}=1-\frac{3h}{2b}+\frac{9h^2+O(a^2)}{4b^2}+O(b^{-3}).
\end{equation}
The first few correction terms to the velocity of the second small sphere are given by
 \begin{equation}
\label{4.6}\frac{U_2}{U_{00}}=1-\frac{3(h+d)}{2b}+\frac{9(h+d)^2+O(a^2)}{4b^2}+O(b^{-3}).
\end{equation}
Hence the velocities of both spheres are less than $U_{00}$, the second slightly more than the first.

It is of interest to note that all the terms shown in Eqs. (4.1)-(4.6) are predicted also by the so-called Rotne-Prager approximation \cite{16}. In this approximation the elements of the mobility matrix are given by single-particle and pair terms of the form
 \begin{equation}
\label{4.7}\vc{\mu}_{ij}=\frac{1}{8\pi\eta}\bigg[\frac{4}{3a_i}\;\delta_{ij}
+\bigg(\frac{\du{I}+\hat{\vc{r}}_{ij}\hat{\vc{r}}_{ij}}{r_{ij}}+\frac{a_i^2+a_j^2}{3}\;\frac{\du{I}-3\hat{\vc{r}}_{ij}\hat{\vc{r}}_{ij}}{r_{ij}^3}\bigg)(1-\delta_{ij})\bigg].
\end{equation}
In the present case $a_0=b$ and $a_i=a$ for $i>0$. The agreement shows that the asymptotic behavior is dominated by the terms $\vc{\mu}_{00}$, $\vc{\mu}_{01}$ and $\vc{\mu}_{10}$ of the mobility matrix, which are well approximated by Rotne and Prager.

\section{\label{V}Discussion}

The explicit expressions we have obtained for the pair and three-body hydrodynamic interactions between a sphere and a number of small particles can be usefully applied in several situations of physical interest. The negative result obtained in Sec. III does not preclude application to the swimming of microorganisms. In particular one can consider linear organisms consisting of a spherical head pushed by a tail of small particles, either in longitudinal or transverse motion. The theory can be based on a general formalism for an assembly of rigid spheres developed by one of us \cite{17}. The detailed analysis involving elastic forces besides hydrodynamic interactions will be presented elsewhere \cite{18}.

A second application of interest concerns the sedimentation of a sphere surrounded by small particles. In Sec. IV we have derived the expressions for the velocity of the sphere and those of the particles in the limit where the sphere radius is large compared to all other distances.

As a third application one could consider the Brownian motion of a sphere in a suspension of small particles. Specifically it would be of interest to study the short-time and long-time diffusion coefficients of the sphere. This would require an analysis based on a generalized Smoluchowski equation. The situation of a macro-ion immersed in an electrolyte solution merits particular study in the present context.\\\\

$\vc{\mathrm{Acknowledgment}}$ M. E.-J. was supported in part by the Polish National Science Centre Grant No. 2011/01/B/ST3/05691 and benefited from the COST Action MP1305.

\end{document}